\documentclass[aps,prl,twocolumn,amsmath,amssymb,showpacs,superscriptaddress]{revtex4-1}
\usepackage{amsfonts,hyperref,times,bm,amssymb}
\usepackage{graphicx}

\begin{document}

\title{Spin resonance in Luttinger liquid with spin-orbit interaction}

\author{O.~A. Tretiakov}
\affiliation{Institute for Materials Research, Tohoku University, Sendai 980-8577, Japan}
\affiliation{Department of Physics \& Astronomy,
	    Texas A\&M University,
            College Station, Texas 77843-4242, USA}
\author{K.~S. Tikhonov}
\affiliation{Department of Physics \& Astronomy,
	    Texas A\&M University,
            College Station, Texas 77843-4242, USA}
\affiliation{Landau Institute for Theoretical Physics, Chernogolovka, Moscow Distr. 142432,
Russian Federation}
\author{V.~L. Pokrovsky}
\affiliation{Department of Physics \& Astronomy,
	    Texas A\&M University,
            College Station, Texas 77843-4242, USA}
\affiliation{Landau Institute for Theoretical Physics, Chernogolovka, Moscow Distr. 142432,
Russian Federation}

\date{March 18, 2013}


\begin{abstract}
 Spin-orbit interaction in quantum wires leads to a spin resonance at
 low temperatures, even in the absence of an external dc magnetic
 field. We study the effect of electron-electron interaction on the
 resonance.  This interaction is strong in quantum wires. We show that
 the electron-electron interaction changes the shape of the resonance
 curve and produces an additional cusp at the plasmon frequency.
 However, except for very strong electron-electron interaction these
 changes are weak since this interaction by itself does not break the
 spin-rotation symmetry that is violated weakly by the spin-orbit
 interaction and external magnetic field.
 \end{abstract}

\maketitle

A recent theoretical work \cite{Pokrovsky2012} has predicted that in
one-dimensional (1D) quantum wires with the spin-orbit interaction
(SOI) \cite{Rashba60} it is possible to observe a relatively sharp
electronic spin resonance (ESR) in terahertz range. An external
magnetic field, which is perpendicular to the internal SOI induced
field, enhances the resonance absorption by several orders of
magnitude leaving the resonance frequency almost unchanged. It occurs
because this field violates a symmetry forbidding the electric dipolar
mechanism of the spin-flip transition. The magnetic field oriented
along the SOI field separates the resonance frequencies of the left-
and right-moving electrons and generates the permanent electric
current and dynamic magnetization.

In the work \cite{Pokrovsky2012} the electrons have been treated as
non-interacting particles.  However, in 1D systems the
electron-electron interaction is known to be strong $V/\epsilon_F\sim
|\ln (na)|/(na)$, where $\epsilon_F$ is the Fermi energy, $n$ is the
1D electron density, $a=\hbar^2\kappa/(me^2)$ is the Bohr's radius in
the material, $m=0.05\, m_e$ is the effective electron mass, and
$\kappa$ is the dielectric constant.  For typical values $n\sim 10^6$
cm$^{-1}$ and $\kappa\sim 20$ the ratio $V/\epsilon_F \simeq 1$.
Therefore, it is important to study the effect of interaction on
electron spin resonance in a quantum wire with the SOI.  This is the
main goal of this paper. The ESR in the Luttinger electron liquid is
the excitation of a spin wave by external ac electromagnetic
field. This resonance would have a simple Lorentzian shape in the
absence of interaction. 

Since the electron-electron interaction is strong, the fermionic
excitations do not exist in 1D systems. They are replaced by bosonic
collective excitations: charge and spin waves. In the framework of
Luttinger model \cite{Gogolin:book, Giamarchi:book} that neglects the
SOI and the deviation of the electronic spectrum near the Fermi points
from the linear behavior, the charge and spin degrees of freedom do
not interact (this is the so-called spin-charge separation). The SOI
separates the Fermi points for different spin projections and makes
possible the resonant spin-flip processes. It was shown that the 
interplay of magnetic field, SOI, and electron-electron interaction 
leads to the formation of spin-density wave state when magnetic field 
is perpendicular to the effective SOI magnetic field \cite{Starykh08}. 
In this Letter we assume that the magnetic field has nonzero 
component along the SOI field. Such a field terminates the spin-density 
wave instability \cite{Starykh08} and simultaneously separates the spin 
resonances for left and right movers  \cite{Pokrovsky2012}. The 
Coulomb interaction is  expected to change the shape of the spin 
resonance line from simple Lorentzian to a power-like one which is 
characteristic for the Luttinger liquid \cite{Gogolin:book, Giamarchi:book, Voit1995}.  
The SOI also violates spin-charge separation and thus enables the
excitation of the charge waves at spin reversal. It can be seen as a
weak resonance at a plasmon frequency instead of the spin-wave
frequency. As we show below, both of these effects really take place,
though both are weak for not too strong electron-electron interaction.

We consider a nanowire with a cross-section so small that electrons
fill partially only the lowest band of the transverse motion (one
channel). In this case the Tomonaga-Luttinger model is applied. The
standard Luttinger liquid theory starts from the fermionic Hamiltonian
with the linearized dispersion \cite{Voit1995, Giamarchi:book} to
which we add the Rashba SOI, $H_R$:
\begin{eqnarray} 
H_0 &=& -i v_F \sum_{\sigma}\int\! dx\, (\psi^{\dagger}_{R,\sigma}
\partial_x 
\psi_{R,\sigma} - \psi^{\dagger}_{L,\sigma}\partial_x \psi_{L,\sigma})
\nonumber \\
&&+H_{int}+H_R.
\label{Hamiltonian0}
\end{eqnarray}
Here the $x$-axis is taken along the quantum wire; $\partial_x =
\partial/\partial x$; $v_F$ is the Fermi velocity; $R,L$ labels the
right and left moving fermions; and $\sigma = \uparrow,
\downarrow$ are the spin projections. The interaction part of the
Hamiltonian, $H_{int}$, contains terms $\rho_{R(L)}(q)\rho_{R(L)}(q)$
quadratic in charge densities $\rho_{R(L)}(q)$ of left and right
movers and the terms quadratic in spin densities such that the total
spin is conserved, e.g. ${:\!\psi^{\dagger}_{R,\uparrow}(x)
  \psi_{L,\uparrow}(x)\!:\,:\!
  \psi^{\dagger}_{L,\downarrow}(x^{\prime})
  \psi_{R,\downarrow}(x^{\prime})\! :}$.

With $H_R =0$ the Hamiltonian~(\ref{Hamiltonian0}) has an obvious
$SU(2)$ symmetry of rotations in the spin space.  The term $H_R =
\alpha\!\int\!  \psi^{\dagger} p_x\sigma_z \psi\, dx$ in the
Hamiltonian $H_0$ represents the Rashba SOI \cite{Rashba60} that
splits Fermi momenta of up and down spins so that four Fermi points
$p_{\rho, \sigma}=\rho p_F -\sigma \alpha m$ appear, but it leaves Fermi
velocities unchanged. The Rashba SOI constant $\alpha$ has
dimensionality of velocity and we assume $\alpha\ll v_F$; $\sigma_z$
is the Pauli matrix; $p_F$ is the Fermi momentum at $\alpha =0$; and
$\rho,\sigma =\pm 1$ correspond to right (left) movers and up (down)
spin projections, respectively. The momenta splitting can be removed
by a single-particle unitary transformation $U=\exp (-i \sigma_z
\alpha mx/\hbar)$ which shifts the momenta by $\pm \alpha m$.  After
this transformation the electronic spectrum becomes the same as
without the SOI and the $SU(2)$ symmetry is restored.

\begin{figure}
\includegraphics{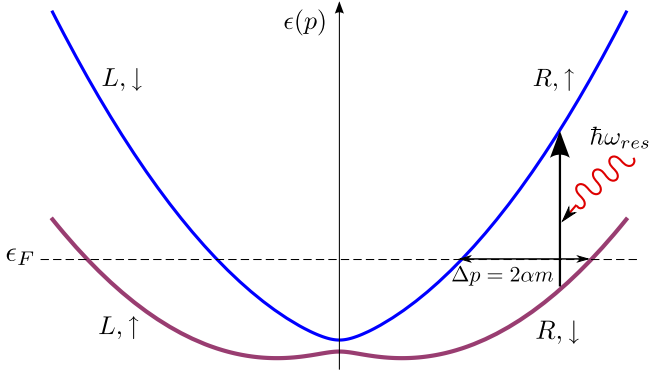} 
\caption{\label{fig:branches} (Color online) The up-spin and down-spin
  branches of the electron spectrum with nonzero Rashba spin-orbit
  interaction $\alpha$ and magnetic field $B_{\perp}$.}
\end{figure}

An external permanent magnetic field breaks this symmetry. It leads to
additional splitting of the Fermi points and to a
difference in Fermi velocities for up and down spins, which cannot be
compensated by this unitary transformation.  We consider in this
Letter only the magnetic field, $B_{\perp}$, perpendicular to the
Rashba field (along $z$-axis) and apply it for definiteness along
$x$-axis. The corresponding Zeeman Hamiltonian reads:
\begin{equation}
\label{Zeeman}
H_Z = -\frac{g\mu_B B_{\perp}}{2}\sum_{\rho,\sigma,\sigma^{\prime}}
\int dx \psi^{\dagger}_{\rho\sigma}(\sigma_x)_{\sigma\sigma^{\prime}}
\psi_{\rho\sigma^{\prime}},
\end{equation} 
where $\mu_B$ is the Bohr magneton and $g$ is the electron g-factor.
Figure~\ref{fig:branches} schematically shows the electron energy as a
function of momentum in the presence of the transverse magnetic field.
We assume that the magnetic field is weak, $g\mu_B B_{\perp}\ll \alpha
p_F$, and further consider it perturbatively.  The residual symmetry
in the perpendicular field is the combined reflection
$p,\sigma\rightarrow -p,-\sigma$. It ensures that the right movers
with the spin projection $\sigma$ along $z$-axis have the same
velocity as the left movers with the same energy and the opposite spin
projection $v_{R,\sigma}=v_{L,-\sigma}$, but $v_{R,\sigma}\neq
v_{R,-\sigma}$. Moroz \textit{et al.}~\cite{MorozPRB1999} have shown
that a velocity difference $\delta v =
v_{R,\uparrow}-v_{R,\downarrow}= v_{L,\downarrow}-v_{L,\uparrow}$
appears also due to the Rashba SOI in the wires of finite width. The
curvature of the bands near Fermi level \cite{Pustilnik06, Khodas07,
  Imambekov09, Pereira2010, SchmidtPRL2010, Adilet2010} can also be
effectively taken into account by the nonzero velocity difference
$\delta v$ on the upper and lower branches of the energy spectrum. The
later effect has a relative value of at most $\sim\alpha/v_F$.

We aim to calculate the absorption power of the resonant ac
electromagnetic field for the spin-flip processes.  The interaction of
electromagnetic field with electrons is described by the Hamiltonian
$H_{em}=-(1/c) \int jA_x dx$, where $j=e\psi^{\dag}(x) \hat{v}\psi
(x)$ is the current, $\hat{v}= \hat{p}/m +\alpha\sigma_z$ is the
velocity operator, and $A_x$ denotes the $x$-component of the vector
potential of the ac field. We use the Coulomb gauge,
$\bm{\nabla}\cdot\mathbf{A}=0$, where the scalar potential is zero,
and therefore the electric field is ${\mathbf
  E}=-(1/c)\partial{\mathbf A}/\partial t$. The part of the electric
current responsible for the spin-flip processes is
\begin{equation}
\label{spin-current}
j_s (x)= e\alpha\psi^{\dag}(x)\sigma_z\psi (x).
\end{equation}
The absorption power of electromagnetic field is determined by the
real part of the conductivity $\sigma_{\omega }$ at the frequency
$\omega$ of the field multiplied by the square of the field's
amplitude $|E_x(\omega )|^2$. We employ the Kubo formula for the
conductivity:
\begin{eqnarray}
\label{Kubo}
\sigma_{\omega} &=& - \frac{1}{\hbar\omega l}
\int_0^l \! dx\int_0^l \! dx^{\prime} \int_{-\infty}^{\infty}\! dt^{\prime}
\theta (t-t^{\prime}) e^{i\omega (t-t^{\prime})}
\nonumber\\
&&\times \langle \left[ j_s(x,t), j_s(x^{\prime}, t^{\prime})\right]\rangle, 
\end{eqnarray}
where $l$ is the length of the wire.  According to
Eq.~(\ref{spin-current}), $j_s(x)$ is proportional to the density of
$z$-component of the spin. Therefore, the spin-flip conductivity
(\ref{Kubo}) can be represented as
\begin{equation}
\sigma_{\omega}=-\frac{4(e\alpha )^2}{\hbar\omega l}
\int_{-\infty}^t\langle\left[S_z(t),S_z(t^{\prime})\right]\rangle
e^{i\omega (t-t^{\prime})}dt^{\prime},
\end{equation}
where $S_z(t)$ is the operator of the total spin projection at the
moment of time $t$. In the absence of magnetic field $B_{\perp}$ the
$z$-component of the total spin is conserved. Therefore,
$\left[S_z(t),S_z(t^{\prime})\right]=0$ and the conductivity
associated with the spin flip is zero.  The violation of this
conservation law at small $B_{\perp}$ in the first-order approximation
of the time-dependent perturbation theory leads to
\begin{equation}
\label{S-change}
\delta S_z(t)=-\frac{i}{\hbar}\int_{-\infty}^t \left[V_I(t^{\prime}),S_z(t^{\prime})
\right] dt^{\prime},
\end{equation}
where $V_I(t)=U_0^{-1}(t)(-g\mu_B B_{\perp}S_x)U_0(t)$ with
$U_0(t)=\exp(-iH_0t/\hbar)$ being the evolution operator in the
absence of magnetic field, and $S_x$ is the projection of the total
spin on the $x$-axis.  It is convenient to write the Rashba
Hamiltonian as a sum over electrons: $H_R=\sum_i \alpha p_i
\sigma_{z,i}$.  The kinetic and interaction energies commute with
$S_x$, and therefore the perturbation operator $V_I(t)$ becomes
\begin{equation*}
 V_I(t)=
 -\frac{g\mu_B B_{\perp}}{2}
  \sum_i \left(\sigma_{x,i} \cos\omega_i t
+\sigma_{y,i} \sin\omega_i t \right),
\end{equation*}
where $\omega_i=2\alpha p_i/\hbar$.  Substituting this expression into
Eq.~(\ref{S-change}), we obtain
\begin{equation}
\delta S_z = \frac{g\mu_B B_{\perp}}
{2\alpha}
\sum_i\frac{1}{p_i}
\left(\sigma_{+,i}e^{-i\omega_i t} +\sigma_{-,i}e^{i\omega_i t}\right),
\label{deltaS1}
\end{equation}
where $\sigma_{\pm}=\sigma_x \pm i\sigma_y$.  Condition $\alpha\ll
v_F$ makes it possible to replace the factor $1/p_i$ in
Eq.~(\ref{deltaS1}) by $\pm 1/p_F$. Then the expression for $\delta
S_z$ becomes proportional to the sum of the operators
$\sigma_{\pm, i}(t)=\sigma_{\pm, i}\exp(\mp i\omega t)$. In terms of
secondary quantized operators it reads (we keep here only right
movers):
\begin{equation}
\label{deltaS}
\delta S_z (t)=  \frac{g\mu_B B_{\perp}}{2\alpha p_F}\int\!\! 
\psi^{\dagger}_{R,\uparrow}(x,t)\psi_{R,\downarrow} (x,t)\, dx +\rm{h.c.}~.
\end{equation}
The unitary transformation $U=\exp (-i\sigma_z\alpha m x /\hbar)$
that puts the split Fermi points together, modifies this equation by
multiplying the integrand by factor $\exp (-2i\alpha mx /\hbar)$.  As
a result, we find for the conductivity~(\ref{Kubo}) associated with
the spin flip \footnote{We retain the same notations $\psi_{R,\sigma}$
  for the transformed fermionic operators.},
\begin{eqnarray}
\sigma_{\omega} &=& - \frac{(e g\mu_B B_{\perp})^2}{\hbar\omega l p_F^2}
\!\!\int_0^l \! dx\int_0^l \! dx^{\prime} \!\!\int_{-\infty}^{\infty}\!\! dt^{\prime}
\theta (t-t^{\prime}) e^{i\omega (t-t^{\prime})}
\nonumber\\
&&\times \langle 
[ \psi^{\dagger}_{R,\uparrow}(x,t)\psi_{R,\downarrow}(x,t), 
\psi^{\dagger}_{R,\downarrow}(x^{\prime}, t^{\prime}) 
\psi_{R,\uparrow}(x^{\prime}, t^{\prime})]\rangle
\nonumber\\
&&\times  e^{-2i\alpha m (x-x^{\prime})/\hbar}.
\label{conductivity}
\end{eqnarray}
An attempt to analyze the spin-flip process in Luttinger liquid 
has been made in  Ref.~\onlinecite{Egger01}. The authors assumed that 
Rashba SOI and longitudinal magnetic field only produce the violation of 
$SU(2)$ invariance and affect the velocity difference $\delta v$, but do not 
separate the Fermi points for spin up and down electrons. The physical origin 
of such a model where the Fermi points are not separated (as it should be for 
the realistic Rashba SOI) was not specified in Ref.~\onlinecite{Egger01}.  

Since the electrons propagating in one direction with the same
velocity strongly interact, the bosonic fields $\phi_c$, $\theta_c$
and $\phi_s$, $\theta_s$ related to the charge and spin density waves,
respectively, give a physically more adequate description of
phenomena. The transformation from fermions to bosons (bosonization)
reads:
\begin{equation}
\psi_{\rho,\sigma}=
U_{\rho,\sigma}
\frac{e^{i\rho k_F x}}{\sqrt{2\pi a_0}} e^{-i[\rho\phi_{c}(x)-\theta_{c}(x)
+\rho\sigma\phi_{s}(x)-\sigma\theta_{s}(x)]/\sqrt{2}},
\label{bosonization}
\end{equation} 
where $U_{\rho,\sigma}$ are the Klein factors which ensure the proper
anticommutation relations between the fermion, and $a_0$ is the
ultraviolet cutoff length. The secondary quantized fermionic
wavefunctions $\psi_{\sigma}$ can be represented by the linear
combinations of right-moving and left-moving fermions
$\psi_{\rho,\sigma}$ with the momenta being close to $\pm k_F$, i.e.,
$\psi_{\sigma} = \psi_{R,\sigma} +\psi_{L,\sigma}$. The advantage of
this model is that the interaction energy becomes quadratic in the
charge and spin density bosonic operators. The density of fermions
becomes linear in bosonic fields $\phi_{c,s}$,
\begin{equation}
\rho_{c,s}(x) =-\frac{\sqrt{2}}{\pi}\partial_{x}\phi_{c,s}(x).
\end{equation}

As we have mentioned, the simplest modification of the fermionic
Hamiltonian produced by the Rashba SOI in the absence of magnetic
field can be removed by the unitary transformation.  At nonzero
magnetic field, the fermionic Hamiltonian $H=H_0+H_Z$ after this
transformation takes the form \cite{MorozPRB2000}:
\begin{eqnarray}
H &=& -i v_1\int\! dx\, (\psi^{\dagger}_{R,\uparrow}\partial_x 
\psi_{R,\uparrow} - \psi^{\dagger}_{L,\downarrow}\partial_x \psi_{L,\downarrow})
\nonumber \\
&&-i v_2 \int\! dx\, (\psi^{\dagger}_{R,\downarrow}\partial_x \psi_{R,\downarrow}
- \psi^{\dagger}_{L,\uparrow}\partial_x \psi_{L,\uparrow}).
\label{SO-fermi}
\end{eqnarray} 
A difference of velocities $\delta v = v_1 -v_2$ can arise due to the
SOI effect in a wire of finite width \cite{
  MorozPRB1999, MorozPRB2000, Pokrovsky2012}, magnetic field, and also
quadratic corrections to the electronic dispersion.  After
bosonization (\ref{bosonization}) Hamiltonian (\ref{SO-fermi}) takes the form
\begin{eqnarray}
H &=& \int\frac{dx}{2\pi}\left[v_{c}K_{c}
\left(\partial_x\theta_{c}\right)^{2}+\frac{v_{c}}{K_{c}}
\left(\partial_x\phi_{c}\right)^{2}+v_{s}K_{s}\left(\partial_x\theta_{s}\right)^{2}
\right.
\nonumber \\
&&\left. +\frac{v_{s}}{K_{s}}\left(\partial_x\phi_{s}\right)^{2}
+\delta v\left(\partial_x\phi_{c}\partial_x \theta_{s}
+\partial_x \phi_{s}\partial_x \theta_{c}\right)\right]
\label{SO-bose}
\end{eqnarray}
where $v_{c}$ ($v_s$) is the velocity of plasmons (spinons). We
have omitted the term $\int \cos[2\sqrt{2}\phi_s (x)] dx/(2\pi)$ as
being irrelevant in the renormalization group procedure for
the repulsive interactions ($K_c <1$) \cite{MorozPRB2000}.

To find the conductivity (\ref{conductivity}) we need to calculate the retarded correlation function $I^R_{\uparrow\downarrow,\downarrow\uparrow} (x,t) =-i\theta(t)\langle
[\psi_{R,\uparrow}^{\dagger}(x,t)\psi_{R,\downarrow}(x,t),
  \psi_{R,\downarrow}^{\dagger} (0,0)\psi_{R,\uparrow} (0,0)] \rangle$
in the ground state of the Hamiltonian (\ref{SO-bose}) with fermionic
operators $\psi_{\rho\sigma}$ given by Eq.~(\ref{bosonization}).   
Since the perturbation theory is developed for time-ordered averages in the 
imaginary time $\tau =-it$, it is instructive to go from $I^R (x,t)$ in the 
Kubo formula to the time-ordered product $I^T_{\uparrow\downarrow,\downarrow\uparrow} (x,\tau) =-\langle T_{\tau}\psi_{R\uparrow}^{\dagger}(x,\tau)\psi_{R\downarrow}
(x,\tau) \psi_{R\downarrow}^{\dagger}
(0,0)\psi_{R\uparrow}(0,0) \rangle$. Applying the Wick theorem, we
obtain in terms of bosonic operators:
\begin{eqnarray}
&& I^T_{\uparrow\downarrow,\downarrow\uparrow} (x,\tau) 
\propto -\frac{e^{g(x,\tau)}}{(2\pi a_0)^2},
\nonumber \\
&&g(x,\tau) = 
\sum_{q,\omega}
[1-e^{i(\omega \tau - qx)}] 
\left\langle Y(q,\omega)Y(-q,-\omega) \right\rangle ,
\label{eq:correlator_b}
\end{eqnarray}
where we introduced $e^{Y(x,\tau)}/(2\pi a_0)= \psi_{R\uparrow}^{\dagger}(x,\tau)\psi_{R\downarrow}
(x,\tau) $ so that $Y(x,\tau)=i\sqrt{2}[\phi_{s}(x,\tau) - \theta_{s}(x,\tau)]$ and $\tau > 0$.
After obtaining  $ I^T_{\uparrow\downarrow,\downarrow\uparrow} (x,\tau)$, it can be converted into retarded correlator using $I^R_{\uparrow\downarrow,\downarrow\uparrow} (t)= i\theta(t)[ I^T_{\uparrow\downarrow,\downarrow\uparrow} (t)- (I^T_{\downarrow\uparrow,\uparrow\downarrow} (-t))^*]$ \cite{Giamarchi:book}.

To find the correlation functions of fields $\phi_{s}$ and
$\theta_{s}$ in Eq.~(\ref{eq:correlator_b}) we use the generating
functional $\mathcal{Z} [\mathbf{J}]$:
\begin{equation}
\mathcal{Z} = 
\int\! 
\mathcal{D}\phi_{i}\mathcal{D}\theta_{i}
\exp\left[ \int\! d\tau\!\! \int\!\! dx
\left(-\frac{1}{2}
\overline{\Phi}M\Phi+\overline{J}\Phi\right)\!\right],
\end{equation}
This expression is written
in a matrix form with 4-vectors of the field $\overline{\Phi}
=(\phi_{c}, \phi_{s},\theta_{c},\theta_{s})$ and "current"
$\overline{J}= (J_{1},J_{2},J_{3},J_{4})$. The $4\times 4$ matrix $M$
describes the system Lagrangian and is presented below.  After the
standard Gaussian integration we find
\begin{equation}
\mathcal{Z} [\mathbf{J}] = \left( \det M \right)^{-1/2} 
\exp\left( \frac{1}{2}\overline{J}M^{-1}J\right).
\end{equation}
The bosonic correlation functions from Eq.~(\ref{eq:correlator_b}) are
represented in terms of the elements of matrix $M$ as
\begin{eqnarray}
&&\left\langle \Phi_{i}(x,\tau)\Phi_{j}(0,0)\right\rangle 
=\left.\frac{\delta^{2}\ln \mathcal{Z}}{\delta J_{i}(x,\tau)
\delta J_{j}(0,0)}\right|_{\mathbf{J}=0}
\nonumber \\
&&=\int \frac{d\omega}{2\pi}\int \frac{dq}{2\pi} 
e^{iqx-i\omega\tau}M^{-1}_{ij}(q,\omega).
\end{eqnarray}
The matrix $M$ is symmetric and has the following nonzero elements
$M_{\phi_{c}\phi_{c}}=v_{c} q^{2}/(\pi K_{c})$,
$M_{\phi_{s}\phi_{s}}=v_{s} q^{2}/(\pi K_{s})$,
$M_{\theta_{c}\theta_{c}}=v_{c}K_{c}q^{2}/\pi$,
$M_{\theta_{s}\theta_{s}}=v_{s}K_{s}q^{2}/\pi$,
$M_{\phi_{c}\theta_{c}}= M_{\phi_{s}\theta_{s}} = iq\omega/\pi$, and
$M_{\phi_{c}\theta_{s}}= M_{\phi_{s}\theta_{c}} =\delta v
q^{2}/(2\pi)$.  With these expressions, $g(x,t)$ in
Eq.~\eqref{eq:correlator_b} takes the form
\begin{eqnarray}
&& g(x,\tau) = 2i\!\int\!\!\!\int\!\! \frac{dq d\omega}{(2\pi)^2}
(1-e^{iqx-i\omega\tau}) [M^{-1}_{\phi_{s}\phi_{s}}(q,\omega)
\nonumber \\
&& +M^{-1}_{\theta_{s}\theta_{s}}(q,\omega)-M^{-1}_{\phi_{s}\theta_{s}}(q,\omega) -M^{-1}_{\theta_{s}\phi_{s}}(q,\omega)] .
\label{exponent}
\end{eqnarray}
At zero SOI ($\alpha = 0$) and magnetic field ($B_{\perp} =0$), the
system has $SU(2)$ symmetry of spin rotation. This symmetry prevents
the renormalization of the interaction constant in the spin channel
and therefore $K_s = 1$ \cite{Adilet2010}. A weak SOI ($\alpha\ll
v_F$) and magnetic field ($B_{\perp}\ll \alpha p_F/\mu_B$) only
slightly violate the $SU(2)$ symmetry \cite{Gritsev05}, so that $K_s -
1\sim(\alpha/v_F)^2$ \cite{Gritsev05}. Therefore, in what follows we
put $K_{s}=1$ up to small corrections of order $\alpha^2$.  Thus, with
this precision up to quadratic in $\delta v$ terms we find
\begin{eqnarray*}
&& M^{-1}_{\phi_{s}\phi_{s}}(q,\omega)+M^{-1}_{\theta_{s}\theta_{s}}(q,\omega)
-M^{-1}_{\phi_{s}\theta_{s}}(q,\omega)-M^{-1}_{\theta_{s}\phi_{s}}(q,\omega)
\\
&& \simeq \frac{2\pi i}{q(\omega +i v_{s}q)}
-(\delta v)^{2} 
\frac{\pi q}{4K_{c}}\frac{\left(K_{c}^{2}
+1\right)v_{c}q +2iK_{c}\omega}{(\omega+iv_{s}q)^{2}
(\omega^{2}+v_{c}^{2}q^{2})}.
\end{eqnarray*}
Performing the integration over frequencies, one finds the correlator
as a function of imaginary time. Because of factor $e^{-i\omega\tau}$
only the poles in the lower half-plane of the complex plane $\omega$
contribute to the integral.  After the integration and analytical
continuation, expression~(\ref{exponent}) turns into a sum of
logarithms of the type $C\ln (x\pm v_{c,s}t)$, where $C$ is a
constant.  Inserting this result in Eq.~(\ref{eq:correlator_b}) we
find the corresponding time-ordered fermionic correlator but in real time. It can be converted into retarded correlation function as $I^R_{\uparrow\downarrow,\downarrow\uparrow} (t)= -2\theta(t) \mathrm{Im}\, I^T_{\uparrow\downarrow,\downarrow\uparrow} (t)$  \footnote{See the supplementary material.}, and using Eq.~(\ref{conductivity}) we obtain
\begin{eqnarray}
\label{conductivity-1}
\!\!\!\!\!\! &&\sigma_{\omega} = {\cal A}\, 
\!\!\int_{-\infty}^{\infty}\!\!\!\!\! dx \!\!\int_{0}^{\infty}\!\!\!\!\! e^{i(\omega t-qx)}
[K( t+i \delta) - K(t-i \delta)] dt,\\
\!\!\!\!\!\! &&K(t) = \frac{1}{(x-v_{c}t)^{\lambda}
(x+v_{c}t)^{\mu}
(x-v_{s}t)^{\nu}},
\label{Kt}
\end{eqnarray} 
where the constant
\begin{equation}
\label{coefficient}
{\cal A}= \frac{\left(e g\mu_B B_{\perp}\right)^2 
a_0^{\lambda +\mu +\nu -2}}{2\pi^2 p_{F}^3\alpha}.
\end{equation}
We recall that the wavevector $q$ in the above integral is equal to
$2\alpha m/\hbar$, cf. Eq.~(\ref{conductivity}). The exact numerical factor ${\cal A}$ is obtained here from the comparison with the
noninteracting result of Ref.~\onlinecite{Pokrovsky2012}, see supplementary material for details.  The integrand in
the integral over $x$ has two singularities in the lower half-plane,
at $x=v_{s}t$ and $x=v_{c}t$.  The expressions for the exponents
$\lambda, \mu$, and $\nu$ are as follows  \footnote{They are presented here up to the quadratic order in small $\delta v/v_F$. There is also a contribution $\sim (x+v_{s}t)^{-\beta}$ in Eq.~(\ref{conductivity-1}), but its exponent $\beta\sim (\delta v)^{4}$ is small and can be neglected to order $(\delta v)^{2}$.}  
\begin{eqnarray}
\lambda & = & \left(\delta v\right)^{2}
\frac{\left(1+K_{c}\right)^{2}}{8K_{c}\left(v_{c}-v_{s}\right)^{2}},\\
\mu & = & \left(\delta v\right)^{2}
\frac{\left(K_{c}-1\right)^{2}}{8K_{c}\left(v_{c} +v_{s}\right)^{2}},\\
\nu & = & 2-\left(\delta v\right)^{2}
\frac{K_{c}v_{c}^{2}+\left(K_{c}^{2}+1\right)v_{c}v_{s}+K_{c}v_{s}^{2}}{2K_{c}
\left(v_{c}^{2}-v_{s}^{2}\right)^{2}}.
\end{eqnarray}
To approximate $\sigma_{\omega}$ close to the spin resonance at
frequency $\omega_{res} = v_{s}q=2\alpha m v_s /\hbar$, we take
$\gamma \leq\mid \omega- \omega_{res}\mid \ll \omega_{res}$ with $\gamma$ being the width of the
resonance which we assume to be small \footnote{The resonance width
  $\gamma$ is mostly due to electron-phonon interaction and can be
  estimated to be $\sim 10^9 - 10^{10}$ s$^{-1}$
  \cite{Pokrovsky2012}. For realistic quantum wires $\gamma\ll \omega_{res} \sim
  10^{12}$ s$^{-1}$.}.  In the limit
$(v_{c}-v_{s})\omega_{res}/(v_s\gamma)\gg 1$, we find \footnote{The
  details of this calculation are presented in the supplementary
  material.}
\begin{eqnarray}
\mathrm{Re}\, \sigma_{\omega} &\simeq& 
{\cal A}
\frac{ q^{\nu-1}}
{(v_{c}-v_{s})^{\lambda}(v_{c}+v_{s})^{\mu}
\Gamma(\nu)} 
\nonumber \\
&&\times  \frac{\gamma}{ \left[ (\omega-\omega_{res})^2 +\gamma^2 \right]^{1
-\frac{\lambda +\mu}{2} }}.
\label{spin_res}
\end{eqnarray}
To evaluate
$\sigma_{\omega}$ close to the other singularity, $\omega=2\alpha m
v_{c}/\hbar$, we use similar approximation and obtain
\begin{eqnarray}
\rm{Re}\, \sigma_{\omega} & \simeq &  
{\cal A}\frac{2\pi\lambda q^{\lambda-1}}{\left(2v_{c}\right)^{\mu}
(v_{c}-v_{s})^{\nu} (2-\mu - \nu)}
\nonumber \\
&&\times \frac{\gamma}{\left[(\omega -v_c q)^2+\gamma^2 \right]^{1-\frac{\mu+\nu}{2}}}.
\label{charge_res}
\end{eqnarray}
The plasmon singularity has a character of a weak cusp that can be
detected only at large enough interaction.

Equations (\ref{spin_res}) and (\ref{charge_res}) are obtained under
the assumption of well separated spinon and plasmon peaks, $(v_c -v_s)q\gg \gamma$ 
\footnote{This result corresponds to the case of strong interaction $g_0$, 
i.e. $\delta v/v_F < g_0$. In the intermediate regime of  weak interaction, 
$\delta v/v_F \gtrsim g_0$, there is also a strong mixing of charge and spin 
channels, however  the entire Luttinger description becomes not suitable since 
it does not take into account the curvature effects which are important  in this case.}. In the
opposite case corresponding to the limit of non-interacting fermions,
the peaks at $\omega=\omega_{res}$ and $\omega=v_{c}q$
merge. According to Eq.~(\ref{conductivity-1}), the combined power of
the merged peaks is $\lambda + \nu=2+ (\delta v)^2(1-K_c)^2/[8K_c(v_c+v_s)^2]$. In
the limit of non-interacting fermions $v_{c}\to v_{s}$ and
$K_{c}\to1$, so that the power becomes $2$ which corresponds to the
Lorentzian shape of the spin resonance \footnote{See the case of noninteracting 
electrons in the supplementary material.}. For small interaction $g_0$
between fermions, $K_{c}\simeq 1-g_0/\pi$, $v_{s}=v_{F}$, and
$v_{c}\simeq v_{F}(1+g_0/\pi)$, so that the power
deviates from $2$ by $\sim (\delta v)^2 g_0^2$. Therefore, in the framework of
perturbation theory the shape of the resonance line near
$\omega_{res}$ deviates slightly from Lorentzian. However,
generally $g_0 = (e^{2}/\kappa\hbar v_{F}) |\ln qa |$ can be of the
order of 1.  For repulsive interactions $0<K_c <1$ and for strong
fermionic interaction $K_c \to 0$. In this case the
results~(\ref{spin_res}) and (\ref{charge_res}) show that the shape of
the absorption line may deviate significantly from Lorentzian at
sufficiently strong interaction. 

In conclusion, we have considered the electron spin-flip resonance
caused by internal SOI field in the framework of Luttinger liquid
theory.  We have shown that the electron interaction incorporated does 
not destroy the resonance.
In this theory it is treated as the excitation of a spin wave with the
uniquely specified wavevector.  We have found that the Luttinger
liquid renormalizations almost do not change the Lorentzian shape of
the resonance line at not too strong interaction. It occurs because
the $SU(2)$ symmetry in the spin channel is only slightly violated by
the spin-orbit interaction and weak dc magnetic field. The same small
parameters ensure that the coupling of the spin flip to the charge
channel and therefore the excitation of the plasma oscillations is
weak at the same conditions. Nevertheless, since the Coulomb
interaction in quantum wires is strong, it may be expected that the
deviation from Lorentzian shape of the resonance can be observed
experimentally.  In this work we considered a simplified model of the
SOI whose only effect is the appearance of the difference between
Fermi velocities of up and down spins. The exact consideration of the
quadratic part of the dispersion is still an open problem.

We are indebted to M. Khodas for numerous valuable discussions. We
thank A.~M. Finkel'stein, L.~I. Glazman, O.A. Starykh, and A.~M. Tsvelik for useful
comments. This work has been supported by the DOE under the grant
DE-FG02-06ER46278. O. A. T. also acknowledges support  by the
Grants-in-Aid for Scientiﬁc Research (No. 25800184 and
No. 25247056) from the Ministry of Education, Culture,
Sports, Science and Technology, Japan (MEXT), from NSF under
Grants No. DMR-0757992, ONR-N000141110780, and from the Welch
Foundation (A-1678). K. S. T. has been supported by NHARP.

\bibliography{1D_spin_orbit}

\newpage

\appendix{\Large Supplementary material for ``Spin resonance in Luttinger liquid with spin-orbit interaction''}

\section{Relation between $I^R_{\uparrow\downarrow,\downarrow\uparrow} (t)$ and $I^T_{\uparrow\downarrow,\downarrow\uparrow} (t)$}

The perturbation theory is valid for time-ordered averages in imaginary time, whereas what we need to calculate is a retarded average $I^R_{\uparrow\downarrow,\downarrow\uparrow} (t)$. Therefore, we need a relationship between $I^R_{BA} (x,t) =-i\theta(t)\langle [B(x,t),
 A(0,0)] \rangle$ and  $I^T_{BA} (x,\tau) =-\langle T_{\tau} B
(x,\tau) A(0,0) \rangle$ for imaginary time $\tau$ where $B(x,t)=\psi_{R,\uparrow}^{\dagger}(x,t)\psi_{R,\downarrow}(x,t)$ and $A(0,0)= \psi_{R,\downarrow}^{\dagger} (0,0)\psi_{R,\uparrow} (0,0)$ are boson-like operators. These two types of averages are related by equality \cite{Giamarchi:book}:
\begin{equation}
I^R_{BA} (t)= i\theta(t) \left[ I^T_{BA} (t)- \left(I^T_{A^{\dagger}B^{\dagger}} (-t)\right)^{*} \right],
\end{equation}
which follows from
\begin{eqnarray}
\!\! I^R_{BA} (t) &=& -i \theta(t) \left[ \left\langle B(t)A(0) \right\rangle - \left\langle A(0)B(t) \right\rangle \right],\\
\!\! I^T_{BA} (t) &=& -\left[ \theta(t) \left\langle B(t)A(0) \right\rangle + \theta(-t) \left\langle A(0)B(t) \right\rangle \right].
\end{eqnarray}
For positive time $t>0$,
\begin{eqnarray}
&&I^T_{BA} = -  \left\langle B(t)A(0) \right\rangle,\\  
&&- \left(I^T_{A^{\dagger}B^{\dagger}} (-t)\right)^{*} = \left\langle B^{\dagger} (0)A^{\dagger} (-t) \right\rangle^{*} ,
\end{eqnarray}
and $ \left\langle B^{\dagger} (0)A^{\dagger} (-t) \right\rangle^{*}
=  \left\langle A (-t)B (0) \right\rangle 
=  \left\langle A (0)B (t) \right\rangle
=  \left\langle B^{\dagger} (t) A^{\dagger} (0) \right\rangle^{*}$. In our case, due to the above definitions of $A$ and $B$, $\left\langle B^{\dagger} (t)A^{\dagger} (0) \right\rangle$ differs from $\left\langle B (t)A (0) \right\rangle$ by changing the spin components $\sigma\to - \sigma$. It is equivalent to $Y(x,t)\to -Y(x,t)$ since
we introduced $e^{Y(x,\tau)}/(2\pi a_0)= \psi_{R\uparrow}^{\dagger}(x,\tau)\psi_{R\downarrow}
(x,\tau) $. However, $Y$ enters in all correlation functions quadratically, see Eq.~(\ref{eq:correlator_b}) in the main part, and we conclude that this transformation does not change the correlator. Then $I^R_{BA} (t)= i\theta (t) [ I^T_{BA} (t)-  \left(I^T_{BA} (t)\right)^{*}]$, and we find
\begin{equation}
I^R_{BA} (t)= -2\theta (t) \mathrm{Im}\, I^T_{BA} (t).
\end{equation}

\section{Evaluation of conductivity in Eqs. (25) and (26)}

To find the absorption power of electromagnetic field we need to
calculate $\mathrm{Re}\left(\sigma_{\omega}\right)$, where the
conductivity is given by Eq.~(\ref{conductivity-1}) in the main text.

\begin{figure}
\includegraphics[width=0.99\columnwidth]{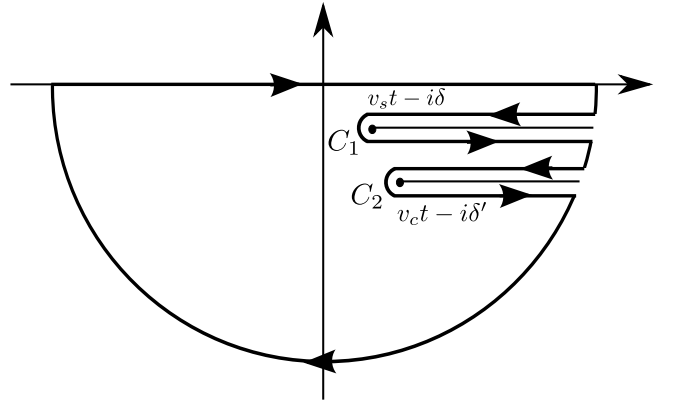} 
\caption{\label{fig:suppl} The integration contour in
  the lower half-plane of the complex variable $x$.}
\end{figure}

The integral in Eq.~(\ref{conductivity-1}) is
\begin{eqnarray}
\label{eq:suppl_eq1}
\!\!\!\!\!\! &&\sigma_{\omega} = {\cal A}\, 
\!\!\int_{-\infty}^{\infty}\!\!\!\!\! dx \!\!\int_{0}^{\infty}\!\!\!\!\! e^{i(\omega t-qx)}
[K( t+i \delta) - K(t-i \delta)] dt,\\
\!\!\!\!\!\! && K(t) = \frac{1}{(x-v_{c}t)^{\lambda}
(x+v_{c}t)^{\mu}
(x-v_{s}t)^{\nu}},
\end{eqnarray} 
where 
\begin{equation*}
\label{coefficient}
{\cal A}= \frac{\left(e g\mu_B B_{\perp}\right)^2 a_0^{\lambda +\mu +\nu -2}}{\pi^2 p_{F}^3\alpha}.
\end{equation*}
Since $q$ is positive, the exponent $e^{-iqx}$ vanishes at large $x$
in the lower half-plane of the complex variable $x$. Therefore, the
integral over real axis $x$ is equal to the sum of two contour
integrals in the lower half-plane of $x$ along contours winding around
two branch cuts shown in Fig.~\ref{fig:suppl}. The contour $C_{1}$
winds around the branch cut from the point $x=v_{s}t-i\delta$ to
$x=+\infty-i\delta$, and the contour $C_{2}$ winds around the branch
cut from $x=v_{c}t-i\delta^{\prime}$ to
$x=+\infty-i\delta^{\prime}$. (We ignore here the potential singularity at $x=-v_{c}t$ which is associated with the inverse processes of spin flip from up to down on the right branch. These processes should be suppressed in the approximation of small excited-state occupation numbers which we employ.) We can estimate the integrals over $x$
around each branch cut separately. The conductivity $\sigma_{\omega}$ has 
two singularities: at $\omega = v_s q$ and at $\omega = v_c q$. As we show 
below, close to the singularity near $\omega = v_s q$ the main contribution to 
the integral comes from contour $C_1$, and the other (plasmon) singularity is 
dominated by the integral over $C_2$.   

First, we estimate the integral $I_1$ over the contour $C_1$. After
the change of variable $u=x-v_{s}t$ the contour that maps $C_1$ in a
complex plane $u$ winds around the branch cut from $u=0$ to
$u=+\infty$ and will be denoted by the same symbol $C_1$. Thus, the
integral $I_1$ can be written as follows:
\begin{equation*}
\label{I1}
I_1=
\int_{C_1}du\frac{e^{-iqu-iv_sqt}}{u^{\nu}\left[u+\left(v_{s}-v_{c}\right)t
\right]^{\lambda}\left[u+\left(v_{s}
+v_{c}\right)t\right]^{\mu}}.
\end{equation*}

We aim to approximate $\sigma_{\omega}$ very close to the resonance at
$\omega_{res}= v_s q$. The closeness is determined by inequalities
relating to the detuning $\delta\omega
=|\omega-\omega_{res}|$:
\begin{equation}\nonumber
\gamma\ll \delta\omega\ll\omega_{res},
\end{equation}
where $\gamma$ is the attenuation rate mainly due to Cherenkov
emission of phonons \cite{Pokrovsky2012}.  Then the time during which
the resonance absorption is accumulated is large enough, $t\sim
1/\delta\omega$. In this limit, $( v_{c}-v_{s})qt\gg 1$, we
approximate the above integral as
\begin{equation}
I_1 =
\frac{e^{-iv_s qt}(-iq)^{\nu-1}}{\left[\left( v_{s}-v_{c}\right) t
\right]^{\lambda}\left[\left(v_{s}+v_{c}\right)t\right]^{\mu}}
\int_{C}z^{-\nu}e^{z}dz,
\end{equation}
where we introduced new variable $z=-iqu$. As a result of this change
of variables, the contour $C_1$ turns into contour $C$ winding around
a branch cut going from $z=0$ to $z=-i\infty$.  The contour $C$ can be
rotated clockwise together with the branch cut until the latter
coincides with the left half of the real axis $z<0$.  Then the contour
integral turns into the Hankel's representation of the inverse Gamma
function and gives $2\pi i/\Gamma(\nu)$. After that we can integrate
over time with the following final result:
\begin{equation}
I_{\omega,q} \simeq {\cal A}
\frac{-2\pi i\Gamma(1-\lambda-\mu)q^{\nu-1}i^{-\nu-\lambda-\mu}}{
\left(v_{s}-v_{c}\right)^{\lambda}\left(v_{s}+v_{c}\right)^{\mu}
\Gamma(\nu)\left(\omega-v_{s}q
+i\gamma\right)^{1-\lambda-\mu}}.
\end{equation}
Then the real part of conductivity near $\omega =
v_sq$ becomes
\begin{eqnarray}
\mathrm{Re}\, \sigma_{\omega}  &\simeq& {\cal A}
\frac{ \Gamma(1-\lambda-\mu) q^{\nu-1}}
{(v_{c}-v_{s})^{\lambda}(v_{c}+v_{s})^{\mu}
\Gamma(\nu)} 
\nonumber \\
&&\times \frac{\gamma}{ \left[ (\omega-\omega_{res})^2 +\gamma^2 \right]^{1
-\frac{\lambda +\mu}{2} }},
\end{eqnarray}
which for small $\lambda$ and $\mu$ is approximated by Eq.~(\ref{spin_res}).
Here we used that $v_{c}\geq v_{s}$. 
The integral $I_2$ over the
contour $C_2$ does not contribute to the singularity at $\omega =
v_sq$ and therefore can be neglected. 

In addition, there is also a small part of $\mathrm{Re}\, \sigma_{\omega}$ which is independent of  $\gamma$:
\begin{equation}
\frac{ \pi^2 {\cal A}\Gamma(1-\lambda-\mu) q^{\nu-1}
(\nu + \lambda +\mu -2)}{(v_{c}-v_{s})^{\lambda}(v_{c}+v_{s})^{\mu}
\Gamma(\nu)|\omega-\omega_{res}|^{1-\lambda -\mu}}.
\end{equation}
It is small in positive parameter $\nu + \lambda +\mu -2 = (\delta v)^2(K_c -1)^2/[4K_c (v_c +v_s)^2]$.
 
Nevertheless, $I_2$ contributes to a
singularity at plasma frequency $\omega = v_c q$. Next we analyze this
singularity.
We consider the integral $I_2$ over contour $C_2$ similarly to what we
did for $I_1$. We change variable in $I_2$ to $u=x-v_{c}t$. The mapped
contour $C_2$ winds around the branch cut from $u=0$ to
$u=+\infty$. As a result of winding around the branch cut we obtain
factor $\left(1-e^{-2\pi i(\lambda-1)}\right)\approx 2\pi i\lambda$
and the integral over $u$ from 0 to infinity:
\begin{equation}
I_2= 2\pi i\lambda\int_{0}^{\infty}du\frac{e^{-iq\left(u+v_{c}t\right)}}{u^{\lambda}\left(u+2v_{c}t\right)^{\mu}\left[u+(v_{c}-v_{s})t\right]^{\nu}}.
\end{equation}
At small detuning $\delta\omega = |\omega -v_c q|$ from the plasma
resonance we expect that similarly to what we observed for $I_1$ the
accumulation time for the resonance absorption is large, $t\sim
1/\delta\omega\gg 1/(v_c q)$, and therefore in the factors $u+2v_c t$,
$u+(v_c -v_s)t$ it is possible to neglect $u\sim 1/q$.  After this
procedure the resulting integral over $t$ diverges at $t=0$. This
divergence however is spurious. It has happened because at small
$t<1/[(v_c-v_s)q]$, the variable $u$ cannot be neglected.  It means
that the integration over $t$ is effectively cut off at $t_0 \sim
1/[(v_c-v_s)q]$.  To estimate the singular part on the background of
nonsingular contribution originated from small $t$, we represent the
exponent $e^{i(\omega -v_c q)t}$ as a sum, $e^{i(\omega -v_c q)t}= [
  e^{i(\omega -v_c q)t} -1] + 1$, and divide the integral over time
into two parts:
\begin{equation*}
\int_{t_0}^{\infty} \frac{\left[e^{i(\omega  -v_c q)t}-1\right]dt}{t^{\mu +\nu}}
+\int_{t_0}^{\infty} \frac{dt}{t^{\mu+\nu}}.
\end{equation*}
The second integral is approximately equal to
$t_0^{1-\mu-\nu}/(\nu+\mu-1)$ and has no singularity. The first
integral converges and can be extended to $t=0$ if $\mu+\nu<2$. This
condition is satisfied in a broad range of not too strong interaction
as it can be readily checked from Eqs.~(22) -- (24). The first
integral after the change of variable $\tau = (\omega - v_c q) t$
turns into
\begin{equation}
\label{auxiliary2}
\left(\omega- v_c q\right)^{\mu+\nu-1}\int_0^{\infty}\frac{\left( e^{i\tau} - 1\right)d\tau}{\tau^{\mu+\nu}}.
\end{equation}
The integral in Eq. (\ref{auxiliary2}) is a large number $\approx i
(2-\mu-\nu)^{-1}$ proportional to $[\delta v/(v_c -v_s)]^{-2}$. The ratio of the
first term to the second has the order of magnitude $[\delta
v/(v_c-v_s)]^{-2}[|\omega-v_c q|/(v_c q)]^{\mu+\nu -1}$. Thus, the nonresonant
contribution is comparable with the resonant one only in a narrow
region close to the resonance $\delta\omega\leq v_c q [\delta
v/(v_c -v_s)]^2$. Combining all the results, we arrive at the expression for the
singularity due to spin-flip processes at the plasmon frequency:
\begin{equation}
I_{\omega,q} \simeq -{\cal A}i^{\lambda -1}q^{\lambda-1} 
\frac{2\pi\lambda\Gamma (1-\lambda)}{2-\mu-\nu}
\frac{\left(\omega -v_c q +i\gamma\right)^{\mu +\nu -1}}{(2v_c)^{\mu}(v_c-v_s)^{\nu}}.
\label{Iomega}
\end{equation}
The calculation of the real part gives the following result:
\begin{eqnarray}
\mathrm{Re}\, I_{\omega,q} &\simeq& {\cal A}\frac{2\pi\lambda 
\Gamma(1-\lambda)q^{\lambda-1}}{\left(2v_{c}\right)^{\mu}(v_{c}-v_{s})^{\nu} (2-\mu - \nu)}
\nonumber \\
&& \times \frac{\gamma}{\left[(\omega -v_c q)^2+\gamma^2 
\right]^{1-\frac{\mu+\nu}{2}}},
\end{eqnarray}
c.f. (\ref{charge_res}) in the main text for $\mathrm{Re}\, \sigma_{\omega}$.
The plasmon singularity has a character of a weak cusp that can be
detected only at large enough interaction.

\section{Case of noninteracting electrons }
\label{nonint_case}

The rate $w$ of the spin flips per one electron found in the work
\cite{Pokrovsky2012} reads:
\begin{equation}
w=\left(\frac{2e\alpha}{\hbar\omega_{res}}\right)^2\left(\frac{g\mu_{B}B}{2\alpha p_F}\right)^2 I_{\omega},
\end{equation}
where $I_{\omega}$ is the spectral density of the driving
electromagnetic field.  It is determined by
\begin{equation}
\overline{\mathbf{E}^{*}(t)\mathbf{E}(t')}= \int_{-\infty}^{\infty} I_{\omega}
e^{i\omega (t-t')}\frac{d\omega}{2\pi}.
\end{equation}
For the monochromatic field with the frequency $\omega_0$ the spectral
density is $I_{\omega}=4\pi\delta (\omega -\omega_0)|{\mathbf
  E}_{\omega}|^2$, where ${\mathbf E}_{\omega}$ is the complex
amplitude of the field. The energy absorption per particle per unit
time is equal to $w\hbar\omega$.  The Luttinger model neglects the
quadratic part of energy dispersion. From this point of view any
particle in the single-occupied range has the same energy of the spin
flip or equivalently the same spin-resonance frequency. To calculate
the energy losses per unit wire length and time it is necessary to
multiply the energy rate per particle by the density of electrons in
the single-occupied ranges of the momentum. The latter is equal to
$n_{sf}=\frac{2\alpha}{v_F}n$, where $n=2p_F/(\pi\hbar)$ is the total
one-dimensional density of electrons. As a result we find the
absorption power per unit wire length:
\begin{equation}
\label{power}
P=w\hbar\omega n_{sf}= \frac{8e^2\left(g\mu_B B\right)^2|\mathbf{E}_{\omega}|^2}{\hbar p_F^2 v_F}\delta (\omega -\omega_{res}).
\end{equation}
The real part of conductivity, $\mathrm{Re}\,\sigma_{\omega}$ for
noninteracting electrons is equal to the power, Eq.~(\ref{power}),
divided by $|\mathbf{E}_{\omega}|^2$:
\begin{equation}
\label{conductivity-ni}
\sigma_{\omega}=  \frac{8e^2\left(g\mu_B B\right)^2}{\hbar p_F^2 v_F}\delta (\omega -\omega_{res}).
\end{equation}
The exact numerical factor for the conductivity of interacting
electrons in Luttinger model is extracted by the matching it with the
conductivity of noninteracting electrons (\ref{conductivity-ni}).

In the case of noninteracting electrons $v_{s}=v_{c}$, $\mu=0$, and
$\lambda+\nu=2$. Substituting it into Eqs.~(\ref{conductivity-1}) -- (\ref{Kt}) of the main text we find ${\cal A}\to {\cal A}|_{\lambda+\mu+\nu=2}$ and Eq.~(\ref{Kt}) simplifies to
\begin{equation}
K(t) = \frac{1}{
\left(x-v_{s}t\right)^{2}}.
\end{equation}
Since $q$ is positive, the exponent $e^{-iqx}$ in  Eq.~(\ref{conductivity-1}) of the main text vanishes at large $x$ in the lower half-plane of the complex variable $x$. Therefore, the
integral over real axis $x$ is equal to the integral in the lower half-plane of $x$. 
Therefore, only $K(t-i\delta)$ has nonzero contribution, and using the residue theorem,
\begin{eqnarray*}
\sigma_{\omega}  &=& {\cal A}
\int_{0}^{\infty}dte^{i\omega(t-i\delta)}
\nonumber \\
&&\times (-2\pi i)\mathrm{Res}_{x=v_{s}(t-i\delta)}
\left[\frac{e^{-iqx}}{\left(x-v_{s}(t-i\delta)\right)^{2}}\right].
\end{eqnarray*}
At a finite attenuation $\gamma$, the conductivity becomes
\begin{equation}
\sigma_{\omega} ={\cal A}
\frac{-2\pi iq}{\omega-v_{s}q+i\gamma}.
\end{equation}
Taking the real part we find in the limit $\gamma\to0$,
\begin{equation}
\mathrm{Re}\,\sigma_{\omega}  = 2\pi^2{\cal A}q\delta\left(\omega-v_{s}q\right).
\end{equation}
Thus, in the noninteracting limit the conductivity of Luttinger liquid
coincides with the conductivity of free electrons provided a correct
choice of the factor $\cal A$.  Note that the cut-off length $a_0$
disappears from the factor ${\cal A}$ in this limit.  At a finite
interaction it enters in a small power $\propto (\delta
v)^2$. Therefore it can be defined only by the order of magnitude.

\end{document}